\documentclass[aip,apl,reprint,superscriptaddress,floatfix,nofootinbib,showkeys]{revtex4-2}
\usepackage[utf8]{inputenc}
\usepackage{amsfonts,amssymb,amsmath}
\usepackage[hidelinks]{hyperref}
\usepackage{graphicx}
\usepackage{color,soul}

\newcommand{\mr}[1]{\mathrm{#1}}
\newcommand{\be}{\begin{equation}}
\newcommand{\ee}{\end{equation}}

\newcommand{\rt}{R_\mr{T}}

\newcommand{\vob}{V_\mathrm{0b}}
\newcommand{\vb}{V_\mathrm{b}}
\newcommand{\ab}{A_\mathrm{b}}
\newcommand{\nng}{n_\mathrm{g}}
\newcommand{\nog}{n_\mathrm{0g}}
\newcommand{\vg}{V_\mathrm{g}}

\newcommand{\ag}{A_\mathrm{g}}
\newcommand{\cg}{C_\mathrm{g}}
\newcommand{\uv}{\,\mr{\mu V}}
\newcommand{\MHz}{\,\mr{MHz}}

\newcommand{\ec}{E_\mathrm{c}}

\begin{document}

\title{Zero-average bias bidirectional single-electron current generation in a hybrid turnstile}

\author{Marco Marín-Suárez}\email{marco.marinsuarez@aalto.fi}
\affiliation{Pico group, QTF Centre of Excellence, Department of Applied Physics, Aalto University, FI-000 76 Aalto, Finland}
\author{Yuri A. Pashkin}
\affiliation{Department of Physics, Lancaster University, Lancaster LA1 4YB, UK}
\author{Joonas T. Peltonen}
\author{Jukka P. Pekola}
\affiliation{Pico group, QTF Centre of Excellence, Department of Applied Physics, Aalto University, FI-000 76 Aalto, Finland}

\begin{abstract}
Hybrid turnstiles have proven to generate accurate single-electron currents.
The usual operation consists of applying a periodic modulation to a capacitively coupled gate electrode and requires a non-zero DC source-drain bias voltage.
Under this operation, a current of the same magnitude and opposite direction can be generated by flipping the polarity of the bias.
Here, we demonstrate that accurate single-electron currents can be generated under zero average bias voltage.
We achieve this by applying an extra periodic modulation with twice the frequency of the gate signal and zero DC level to the source electrode.
This creates a time interval, which is otherwise zero, between the crossings of tunnelling thresholds that enable single-electron tunnelling.
Furthermore, we show that within this operation the current direction can be reversed by only shifting the phase of the source signal.
\end{abstract}


\maketitle

\section{Introduction}

Single-electron turnstiles (SET) are devices that transfer an integer number of electrons between two terminals clocked with a radio-frequency (RF) signal of frequency $f$~\cite{Geerligs1990,Kouwenhoven1991,Blumenthal2007,Pekola2013}.
Consequently, these systems generate, ideally, a current $I=Nef$ with $e$ the elementary charge and $N$ an integer.
These implementations are candidates for a current standard following the revision of the SI (\textit{système international d'unités})~\cite{CGPM2019} in which the ampere has been defined in terms of the electron charge and the unperturbed ground state hyperfine transition frequency of the caesium 133 atom.
Several other proposals have been implemented in order to satisfy metrological requirements for a realization of the ampere, mainly single-electron transport ones~\cite{Pekola2013,Kaestner2015,Yamahata2016}.
Of particular interest have been hybrid single-electron turnstiles composed of superconducting (S) source and drain electrodes separated by insulating (I) barriers from a normal-metal (N) island (SINIS)~\cite{Pekola2007}.
This configuration has been intensively researched as a current standard~\cite{Pekola2007,Maisi2009,Kemppinen2009,Siegle2010,Scherer2012,Zanten2016} among other applications~\cite{Kafanov2009,Knowles2012,Marin-Suarez2020,Marin-Suarez2022}.
So far, the operation of this device has consisted of applying a non-zero source-drain bias defining the current direction.
Additionally, an RF signal was applied to a gate electrode capacitively coupled to the island.
However, recently, it was demonstrated that applying a modulation to the source electrode delivers similar results and even blocks back-tunnelling error events~\cite{Marin-Suarez2022a}.

We show that, by applying RF modulation to the source electrode and following a similar driving strategy as in Ref.~\citenum{Marin-Suarez2022a}, generation of single-electron currents is possible even when the DC bias signal is zero, therefore having zero average bias within one driving period.
Furthermore, we demonstrate the ability to generate current bidirectionally while keeping zero DC bias, by phase-shifting the source signal.
In the past, this has been achieved in silicon quantum dot single-electron pumps, although this required three waveforms for achieving current generation~\cite{Tanttu2016}.
The proposed scheme has similarities with all-metallic electron pumps~\cite{Pothier1991,Keller1996}, which were among the first single-electron transport ampere realizations and quantum capacitance standard~\cite{Keller1999}, in the sense that current can be generated with more than a single waveform at zero bias and flipped by phase-shifting one of such signals.

\section{Experimental Methods}\label{s:methods}

\begin{figure*}
\includegraphics[scale=0.9]{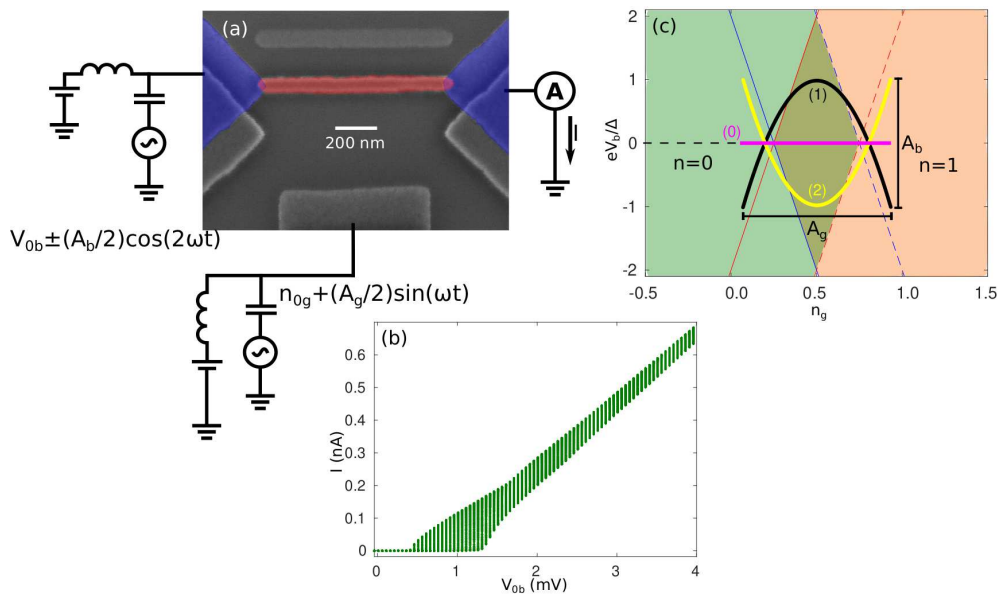}
\caption{(a) Colored electron micrograph with the measurement circuit used.
Red is normal metal, blue is superconductor.
$\vob$, DC bias voltage; $\ab$, peak-to-peak (pp) bias amplitude; $\nog$, DC gate-induced island charge; $\ag$, pp charge-normalized gate amplitude; $I$, generated current; $A$, ammeter.
(b) DC IV curve.
Green vertical lines are measured data within several gate periods.
(c) Stability diagram for a SINIS SET with symmetric junctions zoomed-in to the intersection of the Coulomb diamonds.
Inside the Coulomb diamonds, ideally, no current flows in the DC regime.
Light green diamond corresponds to the island with excess charge $n=0$ and orange diamond with $n=1$.
Lines are tunnelling thresholds, blue (red) corresponds to right (left) junction, continuous (dashed) lines correspond to tunnelling into (out of) the island.
Crossing the line adds or removes one electron to or from the island.
(0) is the conventional gate driving at zero source-drain bias, (1) designates the driving sequence of Eq.~\eqref{eq:protocol} with the plus sign and (2) designates the driving sequence of Eq.~\eqref{eq:protocol} with the minus sign}
\label{f1}
\end{figure*}

\subsection{Fabrication}
We employ a SINIS single-electron transistor such as the one depicted in the false-color electronic micrograph of Fig.~\ref{f1}(a).
We used copper for the normal metal island (red in Fig.~\ref{f1}(a)) and aluminum for the superconducting source and drain leads (blue in Fig.~\ref{f1}(a)).
The device was fabricated using electron beam lithography (EBL) and electron beam double-angle metal evaporation.
The metal films were deposited on top of a silicon oxide coated silicon wafer and through a germanium hard mask~\cite{Meschke2016}.
To form this mask, first the wafer is coated with a layer of poly(methyl methacrylate-methacrylic acid) (P(MMA-MAA)) with thickness of approximately $400\,\mr{nm}$, then we evaporate a $22\,\mr{nm}$ thick layer of germanium on top and finally, we coat with a $50\,\mr{nm}$ thick layer of poly(methyl methacrylate) (PMMA) on top.
Then we use EBL (Vistec EBPG5000+ operating at 100 kV) to expose the PMMA and create the required pattern in the top layer by standard development with a mixture of methylisobutyl-ketone and isopropanol (1 to 3 by weight) after cleaving the wafer into smaller chips for easier handling.
Next, the pattern is transferred to the intermediate germanium layer by reactive ion etching (RIE) in carbon tetrafluoride $\mr{CF_4}$ and then, inside the same chamber, transferred to the bottom P(MMA-MAA) layer by etching with oxygen which also helps to open an undercut profile.
The structure is metallized as follows, first we evaporate $20\,\mr{nm}$ of aluminum at an angle of $15.2^\circ$ forming the leads.
We then perform an \textit{in-situ} oxidation of the surface of this layer in pure oxygen at a pressure of $2.2\,\mr{mbar}$ for 2 minutes.
To form the island we deposit $30\,\mr{nm}$ of copper at an angle of $-14.8^\circ$.
Next, we remove the excess metal and P(MMA-MAA) by soaking the samples in acetone.

\subsection{Measurements}
A chip containing several samples is then cleaved to fit into a custom-made sample carrier.
We incorporated surface mount inductors and capacitors to this carrier to form bias tees for applying AC+DC signals through appropriate input ports to the source and gate electrodes as shown in Fig.~\ref{f1}(a) whereas the drain is connected to a pure  DC line.
The chip is attached to the carrier and connected to the corresponding electrodes by aluminum bonds.
The carrier is then connected to the electrical lines of a custom made dilution refrigerator at a base temperature $\sim 100\,\mr{mK}$ and thermally attached to its mixing chamber.
The DC signals are applied using cryogenic lines composed of resistive twisted pairs from a breakout box at room temperature down to the $1\,\mr{K}$ stage followed by a nearly $1\,\mr m$ long Thermocoax cable down to the mixing chamber.
We apply RF signals through the stainless steel coaxial lines installed between the room temperature and $4.2\,\mr{K}$ flange, then follows a $20\,\mr{dB}$ attenuator.
At the same stage, the line is connected to a feedthrough into the inner vacuum can in which the signal is carried by a superconducting NbTi cable from the $1\,\mr{K}$ flange down to the RF input of the sample carrier.
We add extra attenuation at room temperature, $40\,\mr{dB}$ for the source modulation and $20\,\mr{dB}$ for the gate one.
The signals are generated by programmable DC and waveform generators.
For generating DC signals, the isolated voltage source is used (Stanford Research Sytems SIM928).
For generating RF signals, we use both channels of the 2-channel arbitrary waveform generator (Keysight, model 33522B) therefore ensuring proper synchronization and phase shift between the bias and gate signals.
Current is measured as the output voltage of a transimpedance room temperature amplifier (FEMTO Messtechnik, model LCA-2-10T) which is connected to the drain electrode of the device (ammeter in Fig.~\ref{f1}(a)) via a DC line.
We repeat each measurement of the generated current 15 times and later average it while excluding those repetitions during which a charge offset jump was observed.

\subsection{DC Characteristics and Pumping Method}
The device under test in this work is the same as the one used in Ref.~\citenum{Marin-Suarez2022a}.
The DC characteristics of the system are shown in Fig.~\ref{f1}(b).
We measured the current in this plot by sweeping the DC gate voltage along a couple of charge periods at a fixed DC bias voltage (a single vertical line) at several biases.
The following parameters of the device were determined by calculating the maximum and minimum current in DC with a Markovian model: lead superconducting gap $\Delta =210\,\mr{\mu eV}$, island charging energy $\ec=2.48\Delta$, total normal-state tunnel resistance $\rt=4.53\,\mr{M\Omega}$, ratio between left and right junction resistances $r=0.15$ and Dynes parameter~\cite{Dynes1978} $\eta =3.5\times 10^{-4}$.
See Ref.~\citenum{Marin-Suarez2022a} for an explanation of the model.
Most importantly, it is evident from Fig.~\ref{f1}(b) that for bias voltages $\vob<420\uv$ no current flows irrespective of gate voltage $(\vg)$ and up to certain voltage $\vob\sim 1.46\,\mr{mV}$ current is blocked for certain gate voltages.
This is due to the gap in the superconducting quasiparticle density of states in the leads and the charging energy of the device.
This Coulomb blockade creates a zone in the parameter space of the bias and gate voltage inside which no current flows and the excess charge in the island is stable, this zone is called a Coulomb diamond.
Because of the superconducting gap, the diamonds corresponding to neighbor excess charge states of the island overlap and create a continuous stability region.
We show two of these diamonds in Fig.~\ref{f1}(c) for zero and one excess electron in the island for a device with symmetric junctions.
Notice that these diamonds are bounded by straight lines defined by
\begin{equation}
\Delta =\pm 2\ec\left(n-\nng\pm 0.5\right)\pm eV_\mr{b,L/R}\equiv \delta\epsilon^\pm_\mr{L/R}.
\label{eq:threshold}
\end{equation}
Here $n$ is the charge state of the island, $\nng$ is the charge induced in the island by the gate voltage which relates to it as $\nng=\cg\vg/e$, where $\cg$ is the capacitance between the gate electrode and the island, in our device $\cg=7.28\,\mr{aF}$, $V_\mr{b,L/R}$ is the voltage drop across one junction, if we adopt the convention that positive biasing is from left (L) to right (R) $V_\mr{b,L}=\kappa_\mr L\vb$ and $V_\mr{b,R}=-\kappa_\mr R\vb$, where $\kappa_i$ is the ratio between the junction $i$ capacitance and the total capacitance.
These lines physically correspond to tunnelling thresholds that, when crossed, energetically enable a single tunnelling event into ($+$ in Eq.~\eqref{eq:threshold}) or out of ($-$ sign in Eq.~\eqref{eq:threshold}) the island through the left (L) or right (R) junction, changing its charge state from $n$ to $n\pm 1$.
Notice that these lines are also well defined inside the overlapped stability region.

The turnstile operation consists of  creating a pumping trajectory that crosses these thresholds in a sequence such that in one cycle a fixed number of electrons is transported from one lead to the other.
For example, crossing a line marking the threshold for a tunnelling through one junction into the island and later crossing that for a tunnelling out of the island through the other junction.
This can be achieved by following the path designated as (0) in Fig.~\ref{f1}(c) which is the usual driving sequence that can be achieved by applying any periodic RF waveform to the gate electrode only.
If this is repeated $f$ times per second then one electron is transported in each repetition giving rise to a current $I=ef$~\cite{Pekola2007}.
In Ref.~\citenum{Marin-Suarez2022a} we proposed an alternative driving in which we added to the pure DC bias a periodic modulation such that we could create a parabolic trajectory with negative concavity inside the stability region.
We proved that this new driving scheme also produces single-electron currents and even improves its accuracy.
In the present work we apply a different driving to the SET described by
\begin{equation}
\begin{split}
\nng&=\nog+\dfrac{\ag}{2}\sin{\left(\omega t\right)},\\
\vb&=\pm\dfrac{\ab}{2}\cos{\left(2\omega t\right)},
\end{split}
\label{eq:protocol}
\end{equation}
where $\omega =2\pi f$, $\ab$ is the peak-to-peak (pp) amplitude of the bias signal and $\ag$ is the pp charge-normalized amplitude of the gate signal.
In Eq.~\eqref{eq:protocol}, the $+$ sign generates the path marked as (1) in Fig.~\ref{f1}(c) and the $-$ sign the one designated as (2).
In fact, if we consider time $t$ as a parameter, Eq.~\eqref{eq:protocol} verifies that $\vb=\ab/2\mp4\left(\nng-\nog\right)^2\ab/\ag^2$, that is, parabolas with opposite concavities.
Also, $\nog$ is set to the gate open position at $\vob=0$, that is, $\nog^\mr{open}=0.5$.
Notice that the bias signal averages to zero within one driving period, there is no applied DC level.
In this work, we show that in spite of this, the driving sequence of Eq.~\eqref{eq:protocol} can generate DC single-electron currents whose direction can be changed by switching the phase of the $\vb$ drive signal (the sign in Eq.~\eqref{eq:protocol}).

\section{Results and Discussion}\label{s:results}

\begin{figure*}
\includegraphics[scale=0.78]{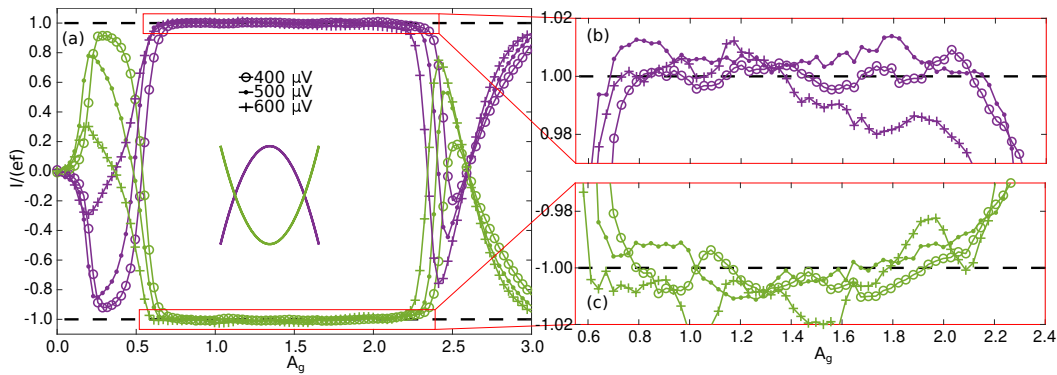}
\caption{(a) Current plateaus produced with the proposed protocol at $f=1\MHz$ against gate signal amplitude, the legend indicates the used pp bias amplitude.
The curves were generated following the paths, depicted at the center of the panel, with the same color as the curves.
(b) Close-up for the current flowing from left to right around $I=ef$ indicated by the dashed line.
(c) Close-up for the current flowing from right to left around $I=-ef$}
\label{f2}
\end{figure*}

The results of this article are shown in Fig.~\ref{f2}.
There we present the current pumped at $f=1\MHz$ against the charge-normalized gate pp amplitude, using the driving methods described by Eq.~\eqref{eq:protocol}.
The curves for different bias pp amplitudes are shown.
It is evident that the use of our driving can generate single-electron currents even though there is no DC bias applied and the bias averages to zero within a driving period.
In Fig.~\ref{f2}(a) the currents generated with both protocols are shown.
Notice that the current generated with protocol (1) generates a positive current and the one generated with protocol (2) is negative, meaning that the direction of the current has been reversed just by a phase-shift of $\pi$ of the bias signal.
Furthermore, notice that these currents form plateaus across a certain interval of $\ag$ values around $I=\pm ef$ and that they are very insensitive to changes in $\ab$ as the overlapping of the different curves suggests, corresponding to different pp bias amplitudes, see Figs.~\ref{f2}(b) and (c).

To have a deeper insight into the single-electron current generation even with zero average bias voltage, we look at the evolution of the quantities $\delta\epsilon^\pm_\mr{L/R}$.
First, we need to understand why no current flows when $\vob=0$ and no modulation is applied to the bias.
Notice from Eq.\eqref{eq:threshold} that, in this case, $\delta\epsilon^+_\mr{L}=\delta\epsilon^+_\mr{R}$ and $\delta\epsilon^-_\mr{L}=\delta\epsilon^-_\mr{R}$, which means that the thresholds for tunnelling through both junctions are crossed simultaneously, no matter whether it is for tunnelling into or out of the island.
This means that an electron is energetically allowed to tunnel through either junction into the island at one instance of the driving period and to tunnel out of the island also through either junction later.
Since tunnelling is stochastic, there are non-zero probabilities of tunnelling through both junctions, but these events will be more frequent through the most transparent one~\cite{Marin-Suarez2022}.
However, when averaging over many periods, as is done in our measurements, all these events cancel each other out resulting in net zero current.
In contrast, when a modulation is applied to the bias, the balance between the tunnelling events through both junctions is lost and $\delta\epsilon^+_\mr{L}\neq\delta\epsilon^+_\mr{R}$ and $\delta\epsilon^-_\mr{L}\neq\delta\epsilon^-_\mr{R}$.
For path (1) we get that $\delta\epsilon^+_\mr{L}=\Delta$ before $\delta\epsilon^+_\mr{R}=\Delta$ with a time difference $\delta t$ and that $\delta\epsilon^-_\mr{R}=\Delta$ before $\delta\epsilon^-_\mr{L}=\Delta$ with the same time difference.
Hence, if $\delta t$ is long enough compared to the tunnelling time, as is the case for $f=1\MHz$, then a DC single-electron current flows from left to right.
In the case of path (2), $\delta\epsilon^+_\mr{R}=\Delta$ before $\delta\epsilon^+_\mr{L}=\Delta$ and $\delta\epsilon^-_\mr{L}=\Delta$ before $\delta\epsilon^-_\mr{R}=\Delta$, generating then a DC current flowing from right to left.

The additional features seen on the current curves of Fig.~\ref{f2}(a) can be explained in terms of the previous description.
For example, the kinks that appear at low $\ag$ with the direction opposite to the current plateau appear because at low amplitudes the path crosses the thresholds in the sequence required for generating this current.
As $\ab$ grows, the $\ag$ interval for this kink narrows as can be deduced from tracing these driving paths in the diagram of Fig.~\ref{f1}(c).
We want to stress that even though the bias DC level is zero, one can engineer many more driving paths with different waveforms such that the thresholds are crossed in a sequence that enables single-electron current generation by applying an appropriate RF signal to the bias.

\section{Conclusions}\label{s:conclusions}

In summary, we have successfully and for the first time, generated single-electron currents in a SINIS SET with zero average bias.
We achieve this by applying a periodic modulation to the source electrode of the device in addition to the conventional gate drive.
This allows to break the balance that exists between the energy cost of tunnelling through either junction when the DC bias is zero and no modulation is applied.
This novel operation regime shows robustness in current accuracy compared to $\lvert I\rvert=ef$ across different bias modulation amplitudes.
Remarkably, this driving allows a current reversal just by shifting the biasing signal phase by $\pi$.
The results are explained in terms of which energy threshold is crossed first, this is readily tuned by the phase of the bias RF signal.

\section*{Acknowledgments}
M.M.-S., J.T.P. and J.P.P. acknowledge support from Academy of Finland under grant number 312057.
Yu.A.P. acknowledges support from the QSHS project ST/T006102/1 funded by STFC.

\section*{Statements and Declarations}
\textbf{Conflict of interest}. The authors declare no competing interests.

\section*{Data Availability}
Data needed for producing the figures and drawing the conclusions of this work are available from the corresponding author upon reasonable request.

\end{document}